\documentstyle[aps,amstex,epsfig,psfig]{revtex}
\begin{document}
\draft
\author{M.B. Z\"olfl$^1$, I.A. Nekrasov$^2$, Th. Pruschke$^1$, V.I. Anisimov$^2$, and J. Keller$^1$}
\address{$^1$Institut f\"ur Theoretische Physik I, Universit\"at
Regensburg, Universit\"atsstr. 31, 93053 Regensburg, Germany\\
$^2$Institute for Metal Physics, 620219 Yekaterinburg GSP-170, Russia}
\date{\today}
\title{The spectral and magnetic properties of $\alpha$- and $\gamma$-Ce
from the Dynamical Mean-Field Theory and Local Density Approximation}
\maketitle
\begin{abstract}
We have calculated ground state properties and excitation
spectra for Ce metal with the {\it ab initio} computational
scheme combining local density approximation and dynamical mean-field
theory (LDA+DMFT). We considered all electronic states, 
i.e. correlated $f$-states and non-correlated $s$-, $p$- and $d$-states. 
The strong local correlations (Coulomb interaction) among the $f$-states 
lead to typical many-body resonances in the partial $f$-density, 
such as lower and upper Hubbard band. Additionally the well known Kondo 
resonance is observed. The $s$-, $p$- and $d$-densities show small  
to mediate renormalization effects due to hybridization. 
We observe different Kondo temperatures for $\alpha$- and $\gamma$-Ce 
($T_{K,\alpha}\approx 1000~K$ and $T_{K,\gamma}\approx 30~K$), 
due to strong volume dependence of the effective hybridization strength for 
the localized $f$-electrons. 
Finally we compare our results
with a variety of experimental data, i.e. from photoemission spectroscopy 
(PES), inverse photoemission spectroscopy (BIS), 
resonant inverse photoemission spectroscopy (RIPES) and 
magnetic susceptibility measurements.    
\end{abstract}

Ce metal is the simplest lanthanide compound with only one atom
in a face centered cubic (fcc) crystal structure and a relatively
small set of relevant electronic states derived from 
$s$-, $p$- $d$- and $f$-orbitals of Ce.
It shows an unique isostructural (fcc to fcc) $\alpha \rightarrow \gamma$
phase transition with increasing temperature. The high-temperature $\gamma$
phase has 15\% larger volume and displays a Curie-Weiss-like temperature
dependence of the magnetic susceptibility signaling the existence of local
magnetic moments while the $\alpha$-phase has a Pauli-like temperature independent
paramagnetism~\cite{exp}. 

While many different models were proposed to describe
this system (for a review see~\cite{old}), the most relevant seems 
to be the periodic Anderson model. Studies based
on the single impurity Anderson model~\cite{Liu} with a hybridization
function obtained from LDA band structure calculations were
rather successful in reproducing Kondo scales and spectra 
for $\alpha$- and $\gamma$-Ce. However, an empirical renormalization
of the hybridization function and  position of the impurity level
were needed for satisfactory agreement between calculated and experimental 
spectra.

Due to the recent development of the Dynamical Mean-Field Theory~\cite{DMFT}
a more realistic treatment of Ce is now possible. 
In contrast to the Hubbard model (degenerate and non-degenerate), where 
hybridization occurs only between correlated
$d$- or $f$-orbitals, Ce is much more complicated.
The direct $f$-$f$ hybridization is of the same order of magnitude as
the hybridization of $f$-orbitals with the delocalized $spd$-states.
Thus in order to describe Ce one even has to go beyond the periodic
Anderson model, where only hybridization of the correlated $f$-orbitals
with the delocalized states is included. In order to address this
problem we used the most general procedure for calculating the Green
function using a full basis set ($s,p,d,f$) Hamiltonian with the integration
over Brillouin zone in k-space.

Recently a LDA+DMFT approach was proposed, with different methods
to solve the DMFT equations: IPT~\cite{poter97,Kajueter,lichten98}, 
NCA~\cite{Zoelfl00,NCA}
and QMC~\cite{Nekrasov00,Held00,rozenberg}. 
All these three methods were used to investigate
La$_{1-x}$Sr$_{x}$TiO$_{3}$~\cite{poter97,Zoelfl00,Nekrasov00}. 
The same strategy was formulated by Lichtenstein and Katsnelson~\cite{lichten98}
as one of their LDA++ approaches. 
Lichtenstein and Katsnelson applied LDA+DMFT(IPT)~\cite{Kats98}, and were 
the first to use LDA+DMFT(QMC)~\cite{kats99} to investigate the spectral 
properties of iron. Liebsch and Lichtenstein also applied LDA+DMFT(QMC) to
calculate the photoemission spectrum of Sr$_{2}$RuO$_{4}$~\cite{liebsch00}.

Here we present results obtained within LDA+DMFT(NCA)~\cite{Zoelfl00,NCA}. 
After convergence of the DMFT selfconsistent loop the hybridization 
function corresponds to fully interacting effective medium for the $f$-shell 
of a Ce ion.
The interacting Green function obtained in our calculation allows us
to compute ground state properties like orbital occupation values,
Kondo temperatures $T_K$, magnetic susceptibility $\chi(0)$, as well as excitation spectra.
Our results show a satisfactory agreement with the experimental data
without using any adjustable parameters.

In the following we will concentrate on a simplified local interaction.
Here, we introduce two distinct Coulomb parameters:
the intra-orbital Coulomb energy $U$ has to be considered in case of a doubly 
occupied orbital, while the inter-orbital Coulomb energy $U^\prime$
applies for example in the case of a doubly occupied $f$-shell
with electrons on $f$-orbitals with different indices. 
Since we neglect any exchange correlations, which is typically of the 
order of one tenth of the Coulomb interaction, 
we chose $U=U^\prime$ in order to fulfill the condition of rotational 
invariance of the local interaction~\cite{MOHM}. 
A more sophisticated interaction term has been already investigated within 
the framework of LDA+DMFT(NCA) scheme~\cite{Zoelfl00}.
We thus arrive at an interaction of the form
\begin{equation}
H^{local}_{corr}= U \sum_m \hat{n}_{m\uparrow}\hat{n}_{m\downarrow}+
\frac{U^\prime}{2}  \sum_{m,m^\prime,\sigma,\sigma^\prime}^{m\ne m^\prime} 
              \hat{n}_{m\sigma} \hat{n}_{m^\prime\sigma^\prime}.
\label{eq1}
\end{equation}
The most important feature of the DMFT is that the proper one-particle 
selfenergy due to the local Coulomb interaction is purely local~\cite{DMFT}.
Thus, we obtain as an expression for the full Green function of the 
interacting system
\begin{equation}
G(z)
=\frac{1}{N_{\vec{k}}} \sum_{\vec{k}} 
(z\overset{\leftrightarrow}{\rm I}
-\overset{\leftrightarrow}{\rm h}(\vec{k})-\Sigma(z)
\overset{\leftrightarrow}{\rm I}_f)^{-1}\;,
\end{equation}
noninteracting one-particle hamiltonian
$\overset{\leftrightarrow}{\rm h}(\vec{k})$ and consequently $G(z)$
will in general be matrices in orbital space,
$\overset{\leftrightarrow}{\rm I}_f$ is the diagonal matrix with
matrix elements equal 1 for f-orbitals and zero for all others,
$\overset{\leftrightarrow}{\rm I}$ is unit matrix.
The $\vec{k}$-summation is done by a standard tetrahedron method
~\cite{poter97}. Within this method one can easily treat hybridization 
effects between correlated and non-correlated states.

As starting point of our calculation we determined the one particle
LDA Hamiltonian with the LMTO method~\cite{LMTO} considering the 
$6s$,$6p$,$5d$ and $4f$-shells. 
The noninteracting one-particle hamiltonian
$\overset{\leftrightarrow}{\rm h}(\vec{k})$ was obtained by
subtracting the Hartree contribution of \eqref{eq1} from the LDA 
results in order to avoid double counting
~\cite{poter97,Zoelfl00,Nekrasov00}.
The value of the Coulomb interaction was calculated by a supercell
method~\cite{ucalc} and found to be $U\approx6~eV$.
The chemical potential was adjusted to conserve 
the number of paricles (4 electrons per site) during the selfconsistent 
LDA+DMFT calculation.
Analizing the partial densities of states one can observe for both $\alpha$- 
and $\gamma$-Ce at a temperature of $T=580~K$ 
intermediate size renormalization effects, in particular broadening and 
shifts of structures, for $s$- and $d$-states and only marginal effects 
for the $p$-states. This is a consequence of hybridization,
which is seen by a non-vanishing $s$- or $d$-density at the position
of the $f$-states in the LDA result. 
The $f$-states are strongly renormalized. A lower Hubbard band (LHB) is 
observed at the position of the corrected (double counting) LDA $f$-state 
(at about $-3~eV$). The upper Hubbard band (UHB) is situated at about $4.5~eV$ and
describes an excitation of a doubly occupied $f$-state. 
At the Fermi level one observes a Kondo resonance (KR) for $\alpha$-Ce, 
which can be described by a singlet formation between an unpaired 
$f$-electron and the surounding conduction electrons.
The $d$-density for $\alpha$-Ce shows an onset of a hybridization gap, 
which is well known in model calculations for the periodic Anderson
model~\cite{periodicAnderson}
and is a consequence of the formation of the singlet state between the
unpaired $f$-spins and the conduction electrons. 
For $\gamma$-Ce one observes only the onset 
of a KR as a consequence of the smaller 
$T_K$ compared to the $\alpha$-phase and thus no
hybridization gap opens in the conduction electron density.

In Table~\ref{occupation} we show a comparison of our results with 
the results of spectral fits to electron~\cite{Liu} and high-energy neutron
spectroscopy ~\cite{Murani}. The occupation probabilities
are in very good agreement with the experiment, as well as the number
of occupied $f$-states per site, $n_f$. The occupation probabilities $P_0$, $P_1$ and $P_2$
for the states $f^0$, $f^1$ and $f^2$ were calculated from the particular ionic propagators. 
The Kondo temperature for $\alpha$-Ce $T_{K,\alpha}\approx 1000~K$
is roughly given 
by the width of the KR.
Since for $\gamma$-Ce the Abrikosov-Suhl resonance is not yet well developed,
it makes no sense to estimate $T_K$ from the width. Instead, 
for $\gamma$-Ce we estimate the $T_K$ from the ratio of the 
hybridization strength at the Fermi level of both considered materials 
($\Im\{\Delta_\alpha(\varepsilon_F)\}/\Im\{\Delta_\gamma(\varepsilon_F)\}=2$). 
With this relation we obtain $T_{K,\gamma}\approx\frac{1}{30}T_{K,\alpha}$
~\cite{Mueller-Hartmann}. Since the NCA for multi-band models typically 
underestimates $T_K$, it is obvious, that our 
absolute values cannot be expected to match with the experiments.
Nevertheless, the ratio of the Kondo temperatures for the two different
phases should be meaningful and are in good agreement to the experiment
(see Table~\ref{occupation}).
The static susceptibilities $\smash[b]{\chi(0)}$ calculated with the $T_K$ 
as an 
input via $\chi(0)=C\frac{\smash[b]{1-n(f^0)}}{\smash[b]{T_K}}$, naturally have 
the same qualitative character. In this formula $C$ is the Curie constant
 for the 
lowest $4f$-state with $j=\frac{5}{2}$ and $n(f^0)$ is the occupation of 
the $f^0$ states. Again, the ratio of the susceptibilities for the both
phases are in good agreement with experiment (see Table~\ref{occupation}).   
Also in Table~\ref{occupation} the parameter $\Delta_{av}$ represents
the averaged value of the imaginary part of the hybridization
function  $\Im\{\Delta(\omega)\}$ for an energy interval from $-3~eV$ 
to $0~eV$~\protect\cite{Anno}. It is also in reasonable agreement with 
experiment.

For $\alpha$-Ce the evolution of the 
KR is clearly observed down to a temperature of
$T=580~K$, whereas in $\gamma$-Ce the evolution of a quasi-particle resonance 
is strongly suppressed in this temperature regime by a smaller hybridization. 
Nevertheless by decreasing the temperature down to $T=116~K$ the onset
of the many-body resonance can be observed for $\gamma$-Ce, too.
 
A comparison of the imaginary part of the hybridization function for 
both phases leads to the result that a strong 
renormalization takes place in comparison to our pure LDA results,
which are in agreement with Ref.~\cite{Liu}. The total weight of this 
quantity for $\alpha$-Ce is twice as high as the one for the 
$\gamma$-phase. Here it turns out that the consideration of all 
hybridization processes is extremly important in order to get a 
qualitatively good agreement to experimental results. If one uses a model
with correlated $f$-states only, one has to consider one electron per 
per site. This would produce a $f$-density with the Fermi 
energy in between the $f^1$ and $f^2$ charge excitation peak, 
the LHB and UHB respectivly. Since the hybridization 
function is proportional to the density of states one would observe 
only a small hybridization at the Fermi energy. Thus the
additional $s$-, $p$- and $d$-conduction states strongly 
contribute to the hybridization function at the Fermi energy and lead 
therefore to different $T_K$.

\begin{figure}
\centering
\epsfig{file=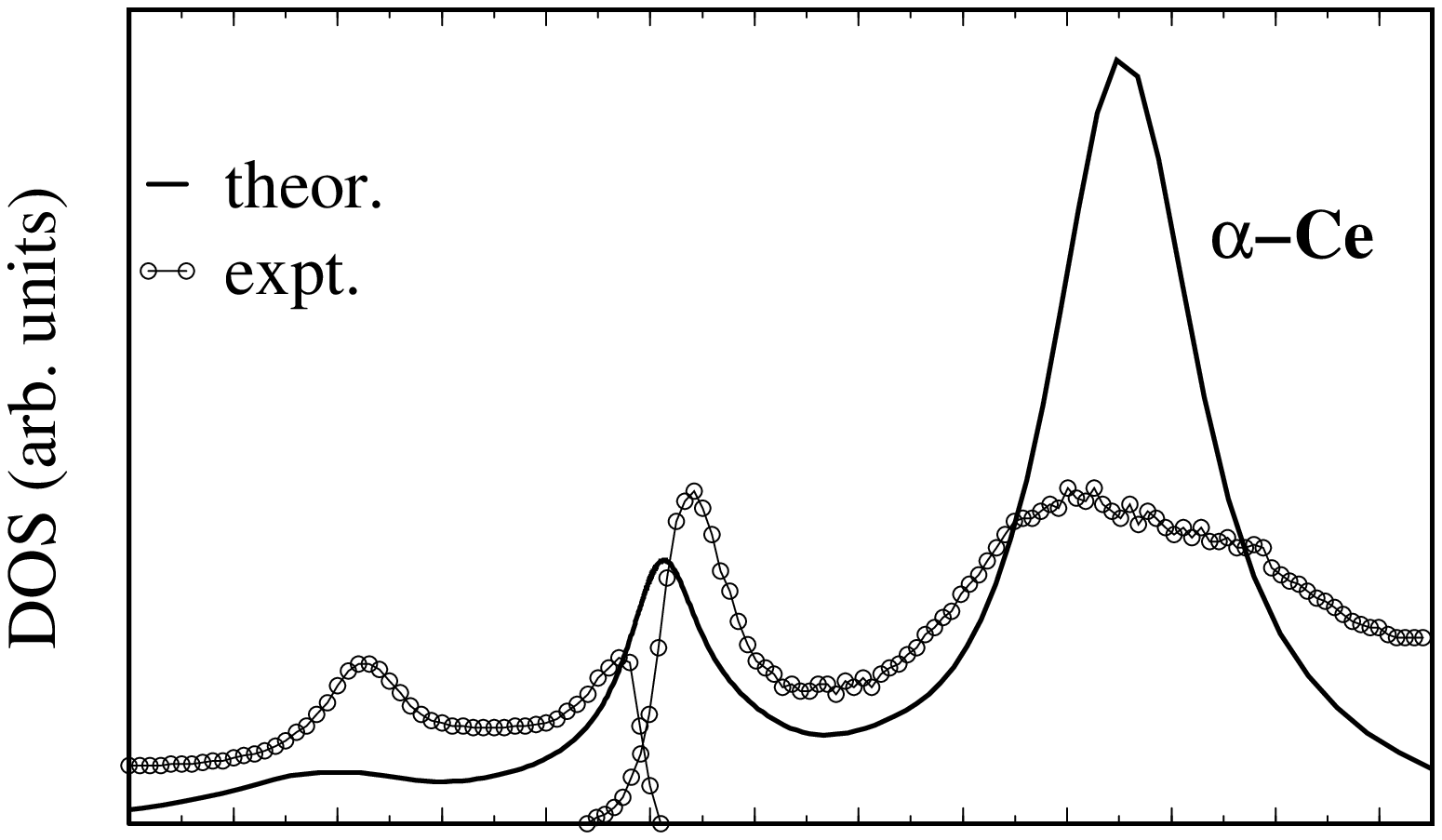,width=0.4\textwidth}\vspace{0.2cm}\\    
\epsfig{file=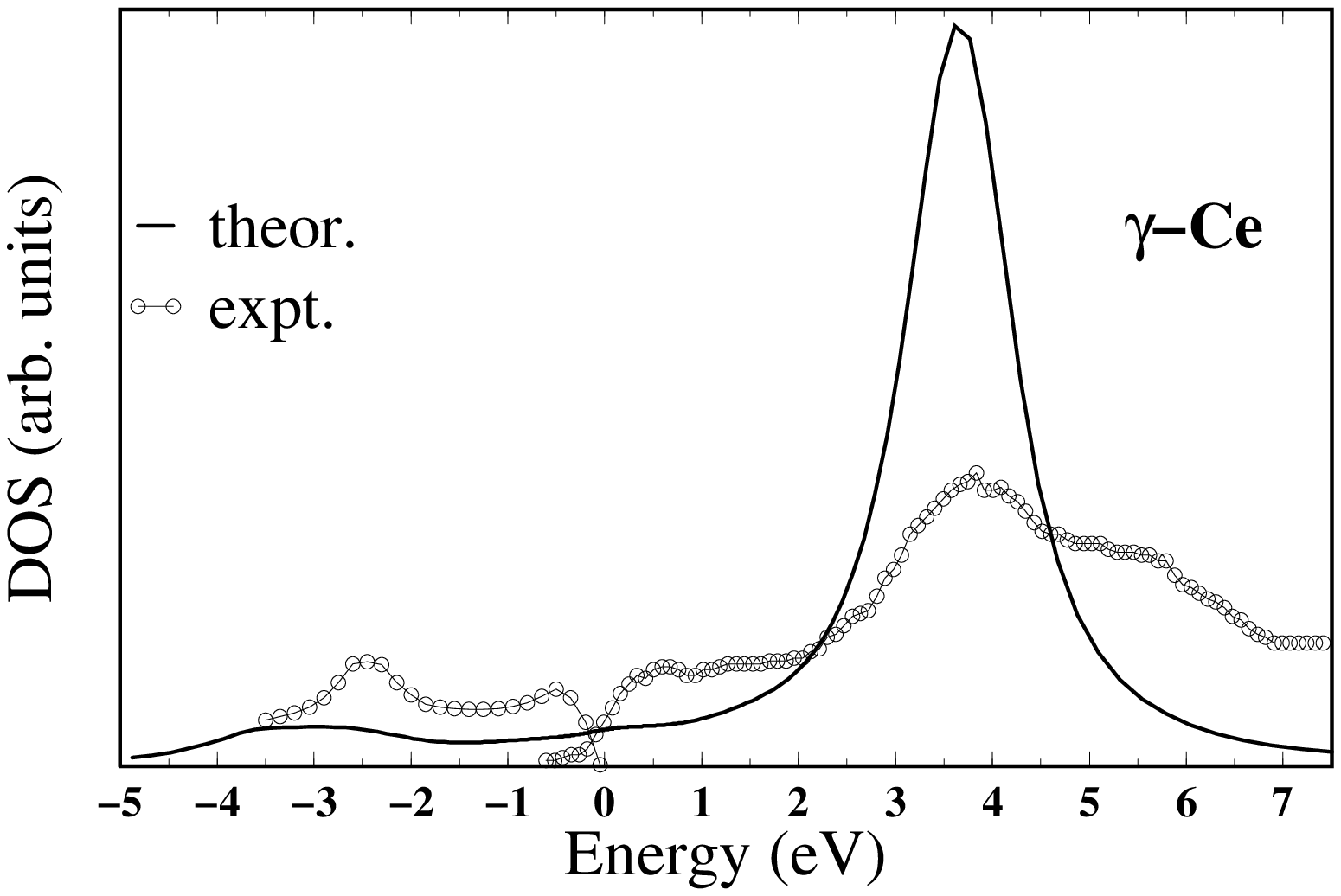,width=0.4\textwidth}  
\caption{Comparison between combined PES~\protect\cite{PES_Wieliczka}
and BIS~\protect\cite{BIS_Wuilloud} experimental (circles) and theoretical (solid line)
$f$-spectra for $\alpha$- (upper part) and $\gamma$-Ce (lower part) at $T=580~K$.
The relative intensities of the BIS and PES
portions are roughly for one 4$f$ electron. The experimental and theoretical spectra
were normalized and the theoretical curve was broadened with resolution width 
of $0.4~eV$.}   
\label{fig:1}
\end{figure}

In the PES data for $\alpha$-Ce in upper part of Fig.~\ref{fig:1} the observed peaks 
are identified as $f$-contributions to the density of states by a cross 
section argument using different photon energies~\cite{PES_Wieliczka}. 
Thus we compare the experiment with the calculated partial density 
for $f$-states.
The theoretical $f$-spectrum shows a LHB which is also seen
in the experiment. 

The BIS spectrum for $\alpha$-Ce shows a main structure between $3~eV$ and $7~eV$, 
which is attributed to $4f^2$ final state multiplets. In the calculated 
spectrum all excitations to $4f^2$ states are described by the featureless
UHB. As a consequence of the simplified interaction model 
all doubly occupied states are degenerate. This shortcoming in our 
calculation is responsible for the sharp peaked structure of this feature. 
The neglected exchange interaction would produce a multiplet structure, 
which would be closer to the experiment. 
The experimental peak at about $0.5~eV$ is attributed to two $4f^1$ final
states, which are split by spin-orbit coupling. 
The calculated $f$-spectrum shows a sharp KR 
slightly above the Fermi energy, which is the result 
of the formation of a singlet state between $f$- and 
conduction states.   
We thus suggest that the spectral weight seen in the experiment is a result
of this KR. Since we did not yet include spin-orbit coupling
in our model, we of course cannot observe the mentioned splitting of the 
resonance. However, as it is well known~\cite{Kondo_degbreak}, the 
introduction of such a splitting would eventually split the 
KR. If we used the experimentally determined value of about 
$0.3~eV$ for the spin-orbit splitting \cite{BIS_Wuilloud}, 
the observed resonance of width $0.5~eV$ would indeed occur in the 
calculations.
 
In the lower part of Fig.~\ref{fig:1} a comparison between experiment and our calculation 
for $\gamma$-Ce is shown. The most striking difference between lower and upper figures 
is the absence of the KR in the high temperature phase 
($\gamma$-Ce; transition temperature $141~K$~\cite{exp}) which is in agreement
with our calculations.

\begin{figure}
\centering
\epsfig{file=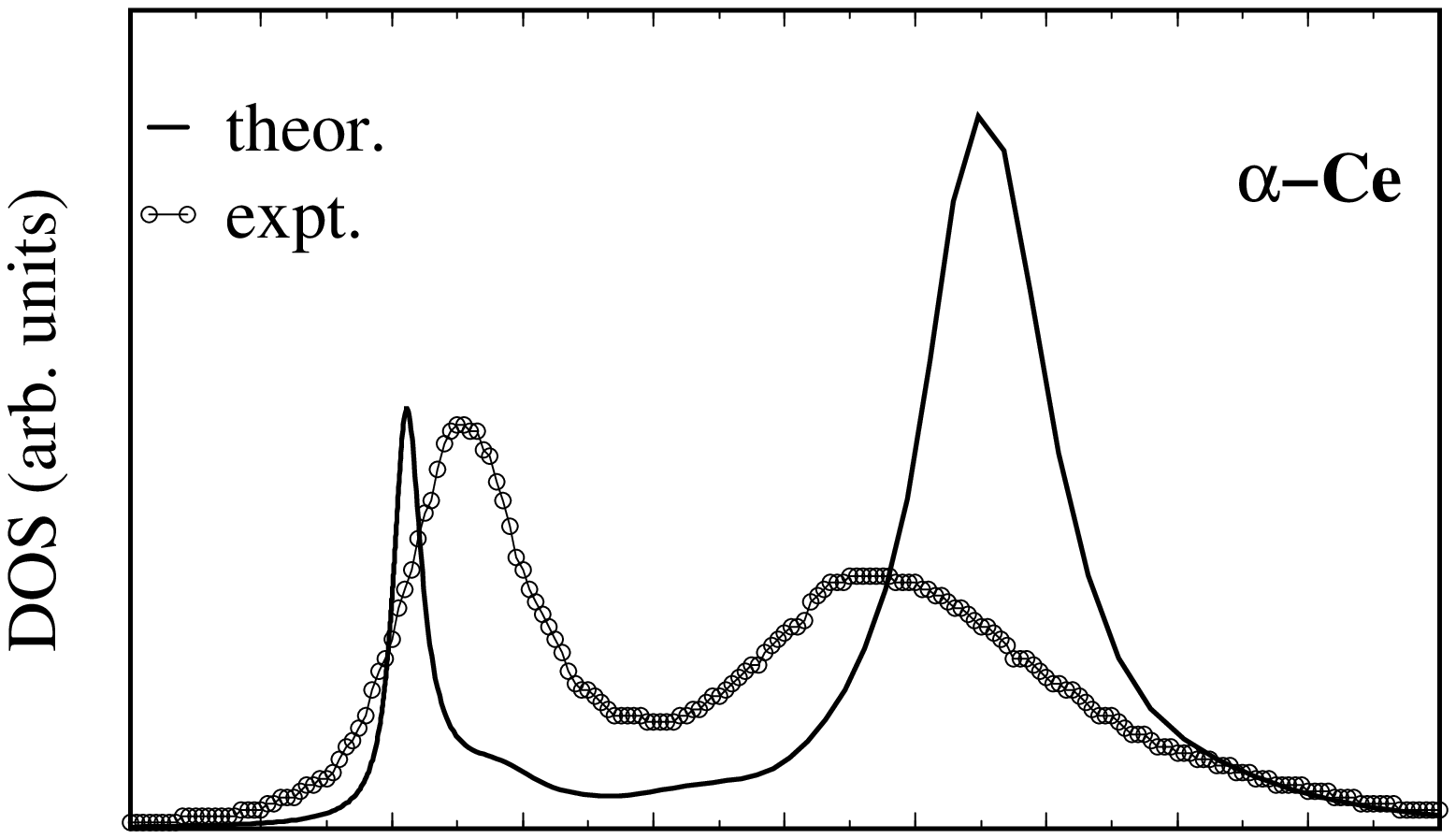,width=0.4\textwidth}\vspace{0.2cm}\\   
\epsfig{file=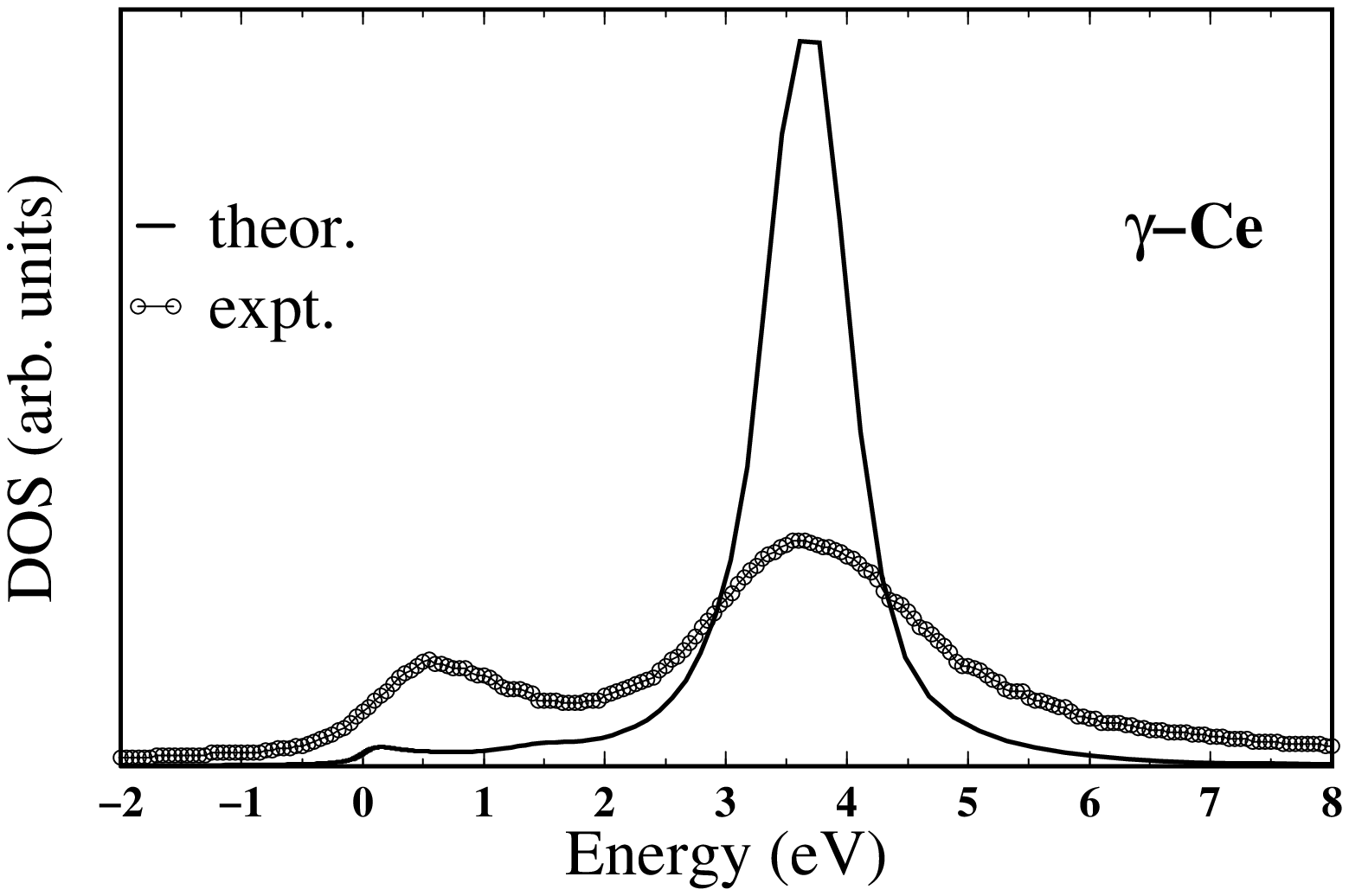,width=0.4\textwidth}  
\caption{Comparison between RIPES~\protect\cite{Grioni}
experimental (circles) and theoretical (solid line) $f$-spectra for $\alpha$- (upper part)
and $\gamma$-Ce (lower part) at $T=580~K$.
The experimental and theoretical 
were normalized and the theoretical curve was broadened with broadening
 coefficient of $0.1~eV$.}    
\label{fig:2}
\end{figure}

In Fig.~\ref{fig:2} our results for the non-occupied states
in the $f$-density are compared with RIPES data~\cite{Grioni}.
The calculated $f$-spectra were
 multiplied by the Fermi-step function and broadened with an 
Lorentzian of the width $0.1~eV$ in order to mimic the experimental resolution 
in the theoretical curves. Here, as above the theoretical overestimation 
of the UHB is a consequence of the simplified local interaction
and thus of the missing multiplet structure of the $4f^2$-final states.
The main feature of the experimental spectra: strong decreasing
of the intensity ratio for KR and UHB peaks going from
$\alpha$- to $\gamma$-phase, can be also seen for
theoretical curves.

In conclusion we have described a realization of a combination
of density-functional theory in the local density approximation and the 
dynamical mean field theory to obtain a first-principles computational
scheme for Heavy-Fermion systems. The scheme was set up 
for the first time with a combination of correlated and non-correlated 
states in order to introduce the important effect of hybridization between 
$s$-, $p$-, $d$-states and strongly correlated $f$-states. 
The solution of the DMFT equations was done by using the Non-Crossing 
approximation. We calculated the one-particle spectra for $\alpha$- and $\gamma$-Ce and 
found Kondo temperature values ($T_{K,\alpha}\approx 1000~K$ and 
$T_{K,\gamma}\approx 30~K$), which explain the experimental results.
 
We observe quite resonable results 
concerning occupation probabilities $P_0$, $P_1$, $P_2$ and the number
of $f$-electrons per site $n_f$.
The ratio of $T_K$ and thus the static susceptibilities $\chi(0)$ 
values for two phases are in 
fair agreement with the experimental results considering the problems of the 
NCA method. Moreover we found qualitative good agreement with PES, BIS and 
RIPES experiments, i.e. the position of LHB, UHB and the Kondo resonance.\\

This work was partially supported by the DFG grant PR 298/5-1\&2 and
Russian Foundation for Basic Research Grant No. RFFI-98-02-17275.

\newpage 

\newpage
\newpage
\begin{table*}
\caption{Comparison between LDA+DMFT(NCA) calculated parameters for both $\alpha$  and $\gamma$ phases
at $T=580~K$ and experimental values.}
\label{occupation}
\begin{tabular}{|p{3cm}|p{3cm}|p{1.5cm}|p{1.5cm}|p{3cm}|p{1.5cm}|p{1.5cm}|}
\hline   & $\alpha$-Ce LDA+DMFT(NCA) & $\alpha$-Ce \cite{Liu} & $\alpha$-Ce \cite{Murani} 
         & $\gamma$-Ce LDA+DMFT(NCA) & $\gamma$-Ce \cite{Liu} & $\gamma$-Ce \cite{Murani}\\
\hline \centering $P_0$ & 0.126 & 0.1558& & 0.0150 & 0.0426& \\ 
\hline \centering $P_1$ & 0.829 & 0.8079& & 0.9426 & 0.9444& \\ 
\hline \centering $P_2$ & 0.044 & 0.0264& & 0.0423 & 0.0131& \\
\hline \centering $n_f$ & 0.908 & 0.861& 0.8 & 1.014 & 0.971& 1\\
\hline \centering $T_K$,~[K] & 1000 & 945 &1800,2000 & 30 & 95 & 60\\
\hline \centering $\chi(0),~[10^{-3}emu/mol$] 
& 1.08  & 0.70 & 0.53 & 24  & 8.0 & 12 \\
\hline \centering $\Delta_{av}$,~[meV] 
& 86.6 & 66.3  &  & 42.7 & 32.2  & \\
\hline 
\end{tabular}
\end{table*}

\begin{references}
\bibitem{exp} K.A. Gschneidner, Jr., R.O. Elliott, and R.R. McDonald, %
J. Phys. Chem. Solids {\bf 23}, 1191 (1962); %
D.C. Koskimaki and K.A. Gschneidner, Jr., %
Phys. Rev. B {\bf 11,} 4463 (1975);
J.M. Lawrence and R.D. Parks, %
J. Phys. (Paris) Colloq. {\bf 37}, C4-249 (1976).

\bibitem{old}  A. K. McMahan, C. Huscroft, R. T. Scalettar, E. L. Pollock %
J. Comput.-Aided Mater. Des. 5, 131 (1998).

\bibitem{Liu}L.Z. Liu, J.W. Allen, O. Gunnarson, N.E. Christensen, %
O.K. Andersen, Phys. Rev. B {\bf 45,} 8934 (1992).

\bibitem{DMFT}D.~Vollhardt in {\em Correlated Electron Systems}, %
edited by V.~J. Emery, World Scientific, Singapore, 1993, p.~57; %
Th. Pruschke, M. Jarrell, and J. K. Freericks, %
Adv. in Phys. {\bf 44}, 187 (1995); %
A. Georges, G. Kotliar, W. Krauth, and M. J. Rozenberg, %
Rev. Mod. Phys. {\bf 68}, 13 (1996).

\bibitem{poter97}  V. I. Anisimov, A. I. Poteryaev, M. A. Korotin, A. O. Anokhin,
and G. Kotliar, J. Phys. Cond. Matter {\bf 9}, 7359 (1997).

\bibitem{Kajueter} H. Kajueter and G. Kotliar, Int. J. Mod. Phys. {\bf 11}, 729
(1997).

\bibitem{lichten98}  A. I. Lichtenstein and M. I. Katsnelson, Phys. Rev. B 
{\bf 57}, 6884 (1998).

\bibitem{Zoelfl00}  M. B. Z\"{o}lfl, Th. Pruschke, J. Keller, A. I. Poteryaev,
I. A. Nekrasov, and V. I. Anisimov, Phys. Rev. B {\bf 61}, 12810 (2000).

\bibitem{NCA}H. Keiter, J.C. Kimbal, Phys. Rev. Lett. {\bf 25,} 672 (1970);
N.E. Bickers, D.L. Cox, J.W. Wilkins, Phys. Rev. B {\bf 36,} 2036 (1987).

\bibitem{Nekrasov00}  I. A. Nekrasov, K. Held, N. Bl\"umer, %
 A. I. Poteryaev,   V. I. Anisimov, and D. Vollhardt, %
Euro Phys. J. B {\bf 18}, 55 (2000).

\bibitem{Held00}K. Held, I. A. Nekrasov, N. Bl\"umer, V. I. Anisimov, and D. Vollhardt, %
preprint cond-mat/0010395.

\bibitem{rozenberg}  M.~J.~Rozenberg,  Phys. Rev. B {\bf 55}, R4855 (1997);
J. E. Han, M. Jarrell, and D. L. Cox, Phys. Rev. B {\bf 58}, R4199 (1998);
K. Held and D. Vollhardt, \newblock { Euro. Phys. J. B \bf 5},
473 (1998).

\bibitem{Kats98}M.I. Katsnelson and A.I. Lichtenstein,
J. Phys. Cond. Matter {\bf 11}, 1037 (1999);

\bibitem{kats99}M.I. Katsnelson and A.I. Lichtenstein, preprint cond-mat/9904428 (1999).

\bibitem{liebsch00}  A. Liebsch and A. Lichtenstein,
Phys. Rev. Lett.{\bf 84}, 1591 (2000).

\bibitem{MOHM} K. Held, D. Vollhard, Euro. Phys. J. B {\bf 5,} 473 (1998). 

\bibitem{LMTO}O.K. Andersen, Phys. Rev. B {\bf 12,} 3060 (1975).

\bibitem{ucalc}V.I. Anisimov, and O. Gunnarsson,
Phys. Rev. B \textbf{43}, 7570 (1991).

\bibitem{periodicAnderson}
Th. Pruschke, R. Bulla, M. Jarrell, Phys. Rev B {\bf 61,} 12799 (2000).

\bibitem{Murani}
A.P. Murani, Z.A. Bowden, A.D. Taylor, R. Osborn, W.G. Marshall,
Phys. Rev B {\bf 48,} 13981 (1993).

\bibitem{Mueller-Hartmann}
E. M\"uller-Hartmann, Z. Phys. B: Condens. Matter {\bf 57,} 281 (1984).

\bibitem{Anno} In our approach we use no compensation factor $\kappa$, 
which was used in~\protect\cite{Liu} in order to consider the effect of a 
self-interaction-correction (SIC).

\bibitem{PES_Wieliczka}
D.M. Wieliczka, C.G. Olson, D.W. Lynch, Phys. Rev. B {\bf 29,} 3028 (1984).

\bibitem{Kondo_degbreak} T. A. Costi, Phys. Rev. Lett. {\bf 85,} 1504 (2000).

\bibitem{BIS_Wuilloud}
E. Wuilloud, H.R. Moser, W.D. Schneider, Y. Baer, Phys. Rev. B {\bf 28,} 7354 (1983).

\bibitem{Grioni}M. Grioni, P. Weibel, D. Malterre, Y. Baer, L. Duo, Phys. Rev. B {\bf 55,} 2056 (1997).

\end{references}
\end{document}